\documentclass[a4paper]{main}
\title{Fast Circular Pattern Matching}
\author{
    Will Solow\inst{1}
\and
    Matthew Barich\inst{2}
\and
   Brendan Mumey\inst{3}
}
\institute{
  Colby College,
  Waterville, Maine, U.S.A\\
  \email{will.solow@colby.edu}
\and
    Western Colorado University,
    Gunnison, Colorado, U.S.A\\
    \email{matthew.barich@western.edu}\\
\and
    Montana State University,
    Bozeman, Montana, U.S.A\\
    \email{brendan.mumey@montana.edu}\\
 }

\authorrunning{Solow, Barich and Mumey}
\titlerunning{Fast Circular Pattern Matching}

\begin{document}

\maketitle

\begin{abstract}
    The Exact Circular Pattern Matching (ECPM) problem consists of reporting every occurrence of a rotation of a pattern $P$ in a text $T$. In many real-world applications, specifically in computational biology, circular rotations are of interest because of their prominence in virus DNA. Thus, given no restrictions on pre-processing time, how quickly all such circular rotation occurrences is of interest to many areas of study. We highlight, to the best of our knowledge, a novel approach to the ECPM problem and present four data structures that accompany this approach, each with their own time-space trade-offs, in addition to experimental results to determine the most computationally feasible data structure.
\end{abstract}

\section{Introduction}
\label{sect:introduction}
\hspace{5mm}Pattern matching has become an increasingly popular topic within computer science due to its presence in many common computer applications including text editors, compilers, and search engines. Pattern matching and the available techniques have also become prominent in the study of computational biology, specifically for sequencing DNA \cite{Azim2015}. 

The pattern matching problem has been studied extensively and can be stated as: Given a pattern $P$ of length $m$, $P=p_1p_2...p_m$, and a text $T$ of length $n$, $T=t_1t_2...t_n$, where all characters in both $P$ and $T$ are drawn from a finite alphabet $\Sigma$ of size $\sigma$, report the indices (if any) of all occurrences of $P$ in $T$. Knuth, Pratt, and Morris developed one of the most well-known pattern matching algorithms, where they match a given pattern into another given string in time proportional to the sum of the lengths of the pattern and string allowing it to run in linear time \cite{Weiner1973}.

Building on the work of Knuth, Pratt, and Morris, Gusfield presented the Z algorithm which allowed for exact pattern matching of a string within the text, with a focus on DNA sequences \cite{gusfield_1997}. As work continued, specifically in the field of computational biology, pattern matching techniques were frequently used when analyzing genomes. In such sub-cases of general pattern matching, as the input text $T$ is generally much longer and relatively static compared to the desired pattern $P$, the option to pre-process the text $T$ became of interest as it could drastically decrease the query time to report all instances of $P$ in on-line searches. With this disregard for pre-processing time, suffix trees were introduced \cite{McCreight}. Suffix trees allow for query time of $O(m)$ using $O(n)$ pre-processing time and space rather than the $O(n)$ query time used in the Knuth-Morris-Pratt algorithm.

A subset of these pattern matching problems in the computational biology field includes finding all circular rotations of a pattern $P$ \cite{ManMachine}. As an example, in the string $banana$, the first rotation of it would be $abanan$ and given a text $T$, we want to report all instances of all seven rotations of the pattern. The problem of Exact Circular Pattern Matching (ECPM) has been studied extensively in \cite{Lin2011} \cite{bhukya_somayajulu_2011} in which the problem is solved in $O(n\log(\sigma))$ time using $O(n)$ space. However, we have not found any prior work solving the ECPM problem using the methods we highlight here. With the disregard for pre-processing time, we can more quickly answer online queries over large texts than possible with previously developed methods.

In addition to the ECPM problem, the approximate circular pattern matching problem (frequently referred to as the $k$-CPM) problem, has been studied rigorously in \cite{Barton2014}\cite{Hirvola2017}. Here, up to $k$ mismatches in the text $T$ are allowed when searching for the pattern $P$ while still being considered a match. This problem is also useful in computational biology as DNA often has inconsequential mismatches that are not of interest. Charalampopoulos presents worst case bounds for two algorithms which are $O(nk)$ and $O(n+\frac nm k^4)$ \cite{CHARALAMPOPOULOS202173}.

\section{Preliminaries}
\label{sect:Preliminaries}
\hspace{5mm}We can state the ECPM problem as follows: given a text $T$ of length $n$, $T=t_1t_2...t_n$ and a pattern $P$ of length $m$, $P=p_1p_2...p_m$, find and report all occurrences of circular rotations, $p_ip_{i+1}...p_mp_1...p_{i-1}$ for $1\leq i\leq m$, using as much pre-processing time as needed. 

This problem can be solved naively by aligning the pattern $P$ and the text $T$ so that the left end of $P$ is aligned with the left end of $T$. Then, each index of $P$ can be compared with $T$ until a mismatch is found, in which case $P$ is shifted one character to the right and the comparisons continue. This process continues until the text $T$ is exhausted. If at any point the pattern $P$ is exhausted with no mismatches found, we know that a match for $P$ has been found and so we report the starting index in $T$ which $P$ was compared to. This method is easy to understand and implement, however, it has a worst-case time complexity of $O(nm)$. To report all circular rotations of $P$ as in the ECPM problem, we could proceed with the above method, repeating the process for every circular rotation for $P$. However, this leads to worst-case time complexity of $O(m^2n)$ - clearly, there are better approaches. 

Given that pre-processing time is not of concern, we utilize a variant of suffix trees, called suffix arrays which were first introduced by Manber and Meyers \cite{Manber1993}. A suffix tree functions by storing every suffix in a tree lexicon-graphically with similar suffixes being on the same branch of the tree. To find a pattern $P$ in a text $T$, it is sufficient to traverse the suffix tree for $T$ following the edges that correspond to the characters in $P$. If $P$ is exhausted, then every child of the node that $P$ ends at is a valid occurrence of $P$ in $T$. A suffix array works similarly but instead stores the index of each suffix in an array. While both suffix arrays and suffix trees required $O(n)$ space, the extra pointers required in a suffix tree implementation can account for gigabytes of space in practice when dealing with large strings \cite{Huo}\cite{Abouelhoda}\cite{sung_2010}. As such, suffix arrays are frequently more popular.

That said, suffix trees can easily return all instances of a pattern in $O(m)$ time by simply traversing the tree by following the edges corresponding to the pattern. While suffix arrays can be augmented using suffix pointers to have an identical worst-case run time, the fastest worst-case search time for a simple suffix array is $O(m\log(n))$ using a binary search algorithm introduced by Nong \cite{Nong}. However, this added factor of $\log(n)$ far outweighs the added space that a suffix tree or suffix pointers require, and in practice is not a noticeable computational bottleneck when pattern matching, even in long texts. When queried with a pattern text $P$ of length $m$, the suffix array returns a start range index and an end range index in which all the occurrences of the pattern reside. This useful property is since all suffixes in the array are lexicon-graphically ordered, which is also the basis of why the binary search method presented in \cite{Nong} performs in $O(m\log n)$ time. The start and end range indices play an integral part in finding all occurrences of a circular pattern in the ECPM problem.

\section{Algorithms}
\label{sect:Algorithms}
To find all circular rotations of a pattern $P$ of length $m$ in a text $T$ of length $n$, observe that $P$ can be broken into two continuous parts, $p_f$ and $p_r$ such that $P=p_fp_r$ and $|p_f|+|p_r|=m$. Now, given every pair of $p_f$ and $p_r$, notice that every circular rotation of $P$ can be written as $rot(P)=p_rp_f$ as the first $|p_f|$ characters of $P$ are moved to the end of the string to create a circular rotation of $P$.

Now, with the above observation in mind, we create two suffix arrays, one with $T$ and one with the reverse of $T$, denoted as $rev(T)$. However, we label the indices of $rev(T)$ with the index of the suffix in $T$ for which it completes the whole string. For example, consider the suffix array and reverse suffix array for the text $T=banana$ (See Table \ref{tab:suffix}).

\begin{table}[h!]
    \centering
    \begin{tabular}{|c | c || c | c|} 
    \hline
        $T$ & & $rev(T)$ & \\
    \hline\hline
        \$ & 6 & \$ & 0 \\ 
    \hline
        a\$ & 5 & ab\$ & 2 \\
    \hline
        ana\$ & 3 & anab\$ & 4 \\
    \hline
        anana\$ & 1 & ananab\$ & 6 \\
    \hline
        banana\$ & 0 & b\$ & 1 \\
    \hline
        na\$ & 4 & nab\$ & 3 \\
    \hline
        nana\$ & 2 & nanab\$ & 5 \\
    \hline
\end{tabular}
\caption{Suffix Trees for $banana$}
\label{tab:suffix}
\end{table}

Notice that the suffix $ana\$$ in $T$ has index 3 which corresponds to the suffix $nab\$$ in $rev(T)$ as the combination of the two completes the whole string $T=banana$. Thus, to search for a circular rotation of $P$, we can search for all occurrences of $p_f$ in the suffix array for $T$ and search for all occurrences of the reverse of $p_r$ in the suffix array for $rev_T$. As we are using suffix arrays, all such occurrences will lie in a continuous range in both $T$ and $rev(T)$. Consequently, if the range $i..j$ in $T$ and the range $k..l$ in $rev(T)$ have any indices in common, then we know that we have found a circular rotation of $P$ in $T$, given how we indexed $rev(T)$. 

By finding an index in common between the two ranges, we have ensured that $p_f$ and $p_r$ occur next to each other in $T$, with $p_r$ coming first. In other words, $p_rp_f$ exists somewhere in $T$, meaning that we have found a circular rotation of $P$. Notice, however, that the index in common between the two ranges is not the index where the circular rotation occurs in $P$, but rather the index in $P$ is the index in common minus the length of $p_r$. 

Ignoring the problem of knowing what indices the two ranges have in common for a moment, observe that to traverse a suffix array to find all occurrences of a pattern of length $m$ takes $O(m\log(n))$ time where $n$ is the length of $T$ \cite{Nong}\cite{Shrestha}. In the worst case, both $p_f$ and $p_r$ have a length on the order of $m$, the length of $P$. So, to traverse the suffix arrays for $T$ and $rev(T)$ and report the ranges of the matches found takes $O(m\log(n))+O(m\log(n))$ which is just $O(m\log(n))$ time. 

However, as we need to compute these ranges for circular rotation of $P$, and as there are $m$ circular rotations, this gives us $(m^2\log^2(n))$ time. In the case of most common applications for pattern matching, the input string $T$ of length $n$ will be many orders of magnitude larger than the pattern $P$ of length $m$. Thus, we observe that this is a very fast algorithm once we solve the problem of computing the indices that the two suffix arrays have in common based on the ranges found. 

To solve the ECPM problem as quickly as possible, we propose the use of an additional data structure to serve as a lookup table. Given two ranges $i..j$ in the suffix array for $T$ and $k..l$ in the suffix array for $rev(T)$, a lookup table will return the indices that the ranges have in common which are exactly the occurrences of a circular rotation of $P$. We propose four different variations of lookup tables, each with their own time-space trade-offs. 

\subsection{Lookup Tables}

In a suffix array, when searching for a pattern $P$, the indices of all occurrences of $P$ will lie in a continuous range in the suffix array as, by definition, a suffix array is sorted in lexicographical order. So, in the suffix array for $T$, all occurrences of $p_f$ will lie in the range $[i..j]$ while all occurrences of the reverse of $p_r$ will lie in the range $[k..l]$ of the suffix array $rev(T)$. 

As a result, the question simplifies to: given two ranges $[i..j]$ and $[k..l]$ in two suffix arrays, do these two ranges have any indices in common when indexing the suffix array $rev(T)$ based on the suffixes of $T$. Notice that by asking how many indices they have in common and by reporting those instances give us exactly the occurrences of a circular rotation of $P$ in $T$. 

As we already noted, there is a time-space trade off in methods that solve this query problem. We outline four methods in order of fastest lookup time to slowest lookup time and most space used to least space used, where $n$ is the length of the text $T$. 

\subsection{Method 1: $O(1)$ Time, $O(n^5)$ Space}
\label{sect:Method 1}
Given two suffix arrays for $T$ and $rev(T)$, each of which has a range of $[1..n]$, we can construct every distinct pair of ranges. Then, for each pair of ranges $[i..j]$ and $[k..l]$, we can iterate through each range comparing the indices and saving the ones that match up into index $i,j,k,l$ in the constructed lookup table. 

Observe that in the range $[1..n]$, there are on the order of $O(n^2)$ ranges, so to store every pair of ranges uses $O(n^4)$ space as we have two suffix arrays each with $O(n^2)$ ranges. In addition, in each range, there can be up to $n$ indices in common (but often much less), giving us our $O(n^5)$ space bound. Already, this method in particularly space-inefficient when processing a three billion character DNA string, as the lookup table will be on the order of gigabytes. 

Notice, however, that with this lookup table construction, given two ranges, we can instantly look up what indices the ranges have in common as it is pre-computed in this lookup table. Thus, we get a lookup time of $O(1)$. 

While pre-processing time to create a lookup table is not directly relevant to the solution of this problem, it is an important consideration as creating such a table could represent a significant computational bottleneck in practical applications, both with construction time and space usage. So, we an algorithm to construct the above lookup table and give its time complexity. 

Simply put, we iterate from $0..n$ in a quad nested for-loop to get every pair of ranges. Note, however, that once we have compared the inner range $k..l$ against the outer range $i..j$, to compute the range $k..(l+1)$ against $i..j$, we only need to compare the index $l+1$ against $i..j$, which saves a significant amount of time. The outer range $i..j$ can be on the order of $n$, and each range in the quad nested for-loop runs from $0..n$, meaning we make on the order of $n^5$ operations. Thus, we get $O(n^5)$ as a worst-case time bound for the construction of this instant lookup table. While still tractable, in practice we will see that this method is very inefficient, even for relatively small strings, and does not scale well to large inputs.
\subsection{Method 2: $O(\log^2n)$ Time, $O(n^3)$ Space}
\label{sect:Method 2}

For this next approach to constructing a lookup table, assume that $T$ is of length $n$ which is a power of 2. Given two suffix arrays for $T$ and $rev(T)$ of length $n$, we build a binary tree array for each suffix array which takes $O(n)$ time. This gives us segments that contain every index in that range for each suffix array which we then can compare point-wise to get every index in common between the suffix array for $T$ and suffix array for $rev(T)$. In a binary tree, the number of segments that we have is on the order of $n$. In fact, when constructing a binary tree it is typical to allocate $4n$ segments as an upper bound. So, to store every pair of segments between the binary arrays will take $8n^2$ space, which is on the order of $O(n^2)$. However, in pair of ranges, there can be at most $n$ indices in common, as when we compare the entire binary tree ranges to each other. In practice, the indices in common will be much smaller than $n$, but this still gives us a worst-case space complexity of $O(n^3)$.

Now, we obtain the worst-case query time bound. Given this paired binary tree of suffix array ranges, we claim that given two pairs of ranges, the query time complexity is $O(\log^2n)$. This hinges on the claim that given a range $[i..j]$ in the suffix array, we can represent it using at most $\log n$ segments in our binary tree. 

Given a range in $1..n$ of length $q$, observe that we need at most two segments of the same size to represent any range. As all ranges are continuous due to the lexicon-graphically ordered nature of a suffix array, this means that the range to be represented will be continuous as well. As such, if there are three segments of the same size used to represent a range, we can replace two of them with a segment of the next larger size containing the same information about indices in common, by nature of this binary tree construction. 

Consequently, for any range, we only need to query to at most two segments of one size to represent any range. In the worst case, we could need two of every sized segment to represent a range. As there are $\log(n)$ segment sizes in a binary tree array, this means that we would need to query at most $2\log(n)$ segments. However, given that we are working with pairs of segments, we must do the same for the binary tree array for $T$ and for $rev(T)$. As such, this gives us a worst case of needing to query to $4\log^2(n)$ segments which leads to a worst-case lookup time complexity of $O(\log^2(n))$, as previously claimed. 

Again, note that time complexity of the construction of a Log lookup table is not of the utmost importance given how we are not concerned with pre-processing time for the direct solution. However, it is again of notable consideration in the application of the problem. We outline the construction of this Log lookup table as follows: After constructing the two binary tree arrays, run through every leaf node and store the indices that are in common at the specific ranges. As there are $n$ leaf nodes in a complete binary tree, this will take $O(n^2)$ time to complete. From there, we can build every range in the binary tree for $T$ in comparison to the leaf nodes of $rev(T)$ by combining the ranges of the children of the node in the binary tree for $T$. As there are $n+1$ internal nodes in the binary tree array for $T$, and $n$ leaf nodes in the binary tree array $rev(T)$, this will again take $O(n^2)$ time to complete. 

Now, we can compare every node in the binary tree array for $rev(T)$ to the internal nodes in the binary tree array for $T$. This works similarly to above, except that we now use the children of the binary tree array for $T$ to construct the segment for the binary tree array for $rev(T)$. As there are on the order of $n$ nodes in $rev(T)$, and $n+1$ internal nodes in the binary tree array for $T$, this again takes $O(n^2)$ time. As such, we have $O(n)+O(n^2)+O(n^2)+O(n^2)=O(n^2)$ time in the worst case to construct a log lookup table for a text $T$ of length $n$. 
\subsection{Method 3: $O(n^{\frac23})$ Time, $O(n^{\frac43})$ Space}
\label{sect:Method 3}

In this next approach, we highlight a root decomposition method to create a lookup table. Given two suffix arrays, one for $T$ and one for $rev(T)$, we decompose each of them into $b$ blocks, each of size $\frac nb$. We then create a lookup table that stores the indices in common between each pair of $b$ blocks. Our goal is to optimize for the value of $b$ by analyzing the lookup time complexity and then compute the space complexity for the root lookup table as well. 

Given two ranges $[i..j]$ and $[k..l]$ in the suffix arrays $T$ and $rev(T)$, each range contains at most $b$ blocks. So, as each block pair contains the indices in common, it will take at most $O(b^2)$ time to query all the full blocks in the worst case (although realistically much less in the average case). However, it is possible to have individual items in the range that are not part of a full block. As each block has size $\frac nb$, the maximum number of items not contained in a full block is $2\frac nb$, for each side of the range of full blocks. Using the inverse table method outlined below, we can query the ranges of length $\frac nb$ in $O(\frac nb)$ time using $O(n)$ space as we store two additional arrays, each of length $n$. Storing this extra space is a worthwhile trade-off as it will make this root decomposition equivalent to the inverse method when the ranges queried are fully inside a block. 

As a result, this gives us a worst case query time of $O(b^2)+O(\frac nb)$. To optimize this, we clearly want $b^2=\frac nb$, or that $b^3=n$. Therefore, $b=n^{\frac 13}$, meaning that we have $n^{\frac13}$ blocks, each of size $n^{\frac23}$. This gives us a query time of $O(n^{\frac23})$ when we substitute this value for $b$ into our original time bound. 

Now, for space, we have $n^{\frac13}$ blocks for each array, meaning that we have $n^{\frac23}$ block pairs which contain the indices that the blocks have in common based on the corresponding suffix arrays. As each block contains at most $n^{\frac23}$ indices by how we constructed each block, this gives us $n^{\frac43}$ as our bound. We also have an additional $O(n)$ space used where we store the two inverse arrays for the inverse method. So, this root decomposition method uses $O(n^{\frac43}) + O(n)$ space, which is on the order of $O(n^{\frac43})$ space. 

To build the required data structures for the root decomposition method, we first build the two inverse arrays which take $O(n)$ time as described in the inverse method. Then, to build the block lookup table, we first must compare every pair of blocks. As there are $b$ blocks, we must compare $b^2$ blocks to each other. Note, however, that each block only contains $\frac nb$ elements, so to compare each pair of blocks, we must make $(\frac nb)^2$ comparisons. So, the total time to build the block lookup table is $O(b^2\cdot (\frac nb)^2)$ which is just $O(n^2)$. Thus, the total build time is $O(n^2)+O(n)$, which is simply $O(n^2)$. 
\subsection{Method 4: $O(n)$ Time, $O(n)$ Space}
\label{sec:Method 4}

In this final method, we highlight a lookup table that makes use of an inverse array. Notice that a suffix array has a well defined inverse, given that the suffixes are a permutation of the indices in a suffix array. Given two suffix arrays for $T$ and $rev(T)$, we create an inverse array from the suffix array of $T$ and an inverse array from the suffix array of $rev(T)$. This uses $O(n)$ additional space given that the text $T$ is of length $n$ and we are simply storing two arrays of length $m$. 

Now, given two ranges $[i..j]$ and $[k..l]$ in the suffix arrays for $T$ and $rev(T)$, we use these ranges to find the indices in common in conjunction with the inverse arrays of $T$ and $rev(T)$. If either of the ranges $[i..j]$ or $[k..l]$ are empty, then we can immediately return no matches, as there are no indices in common. Otherwise, we take the minimum of the two ranges and use it for our query. For each item in the smaller of the two ranges, we look up its value in the inverse array. When the minimum range is $[k..l]$, we look up its value in the inverse array for $T$, and when the minimum range is $[i..j]$ we look up its value in the inverse array for $rev(T)$. If the returned value from the inverse array lies in the other range, then a match was found and it is the value which the was originally queried into the inverse array. 

In the worst case, both $[i..j]$ and $[k..l]$ could be on the order of $n$, meaning that we have to process $n$ such queries through one of the inverse arrays. This gives us a worst-case time bound of $O(n)$. Importantly, observe that this method will always work given that ranges in suffix arrays are continuous, and so we do not risk accidentally finding a match that is not within the corresponding range. While the worst case time complexity is $O(n)$, notice that as we choose the minimum of the two ranges, the average case run time will generally be much better as we almost never query such large ranges except in edge cases.

Observe that to build such an inverse lookup table has a worst-case (and average case) of $O(n)$ time as we run through each suffix array, which is of length $n$, and build the corresponding inverse table. Notice that the general theme between the Instant, Log, Root, and Inverse lookup table methods is that we compromise on pre-processing time and space to get a faster lookup time (see \ref{tab:Methods}). Depending on the application, it may be best compromise on one or the other. This depends on the amount of queries run, but we believe that the inverse lookup table method generally offers the best performance. 

\begin{table}[h!]
    \centering
    \begin{tabular}{| c | c | c | c | c |}
    \hline
    Method & Name & Lookup Time Complexity & Space Complexity & Construction Time \\
    \hline
    1 & Instant & $O(1)$ & $O(n^5)$ & $O(n^5)$ \\
    \hline
    2 & Log & $O(\log^2 n)$ & $O(n^3)$ & $O(n^2)$ \\
    \hline
    3 & Root & $O(n^{\frac 23})$ & $O(n^{\frac 43})$ & $O(n^2)$ \\
    \hline 
    4 & Inverse & $O(n)$ & $O(n)$ & $O(n)$\\
    \hline
\end{tabular}
\caption{Four Methods to create a Lookup Table}
\label{tab:Methods}
\end{table}

\section{Experimental Results}
\label{sect:Ecperimental Results}
\begin{figure}[!h]
\centering
  \includegraphics[width=.75\textwidth]{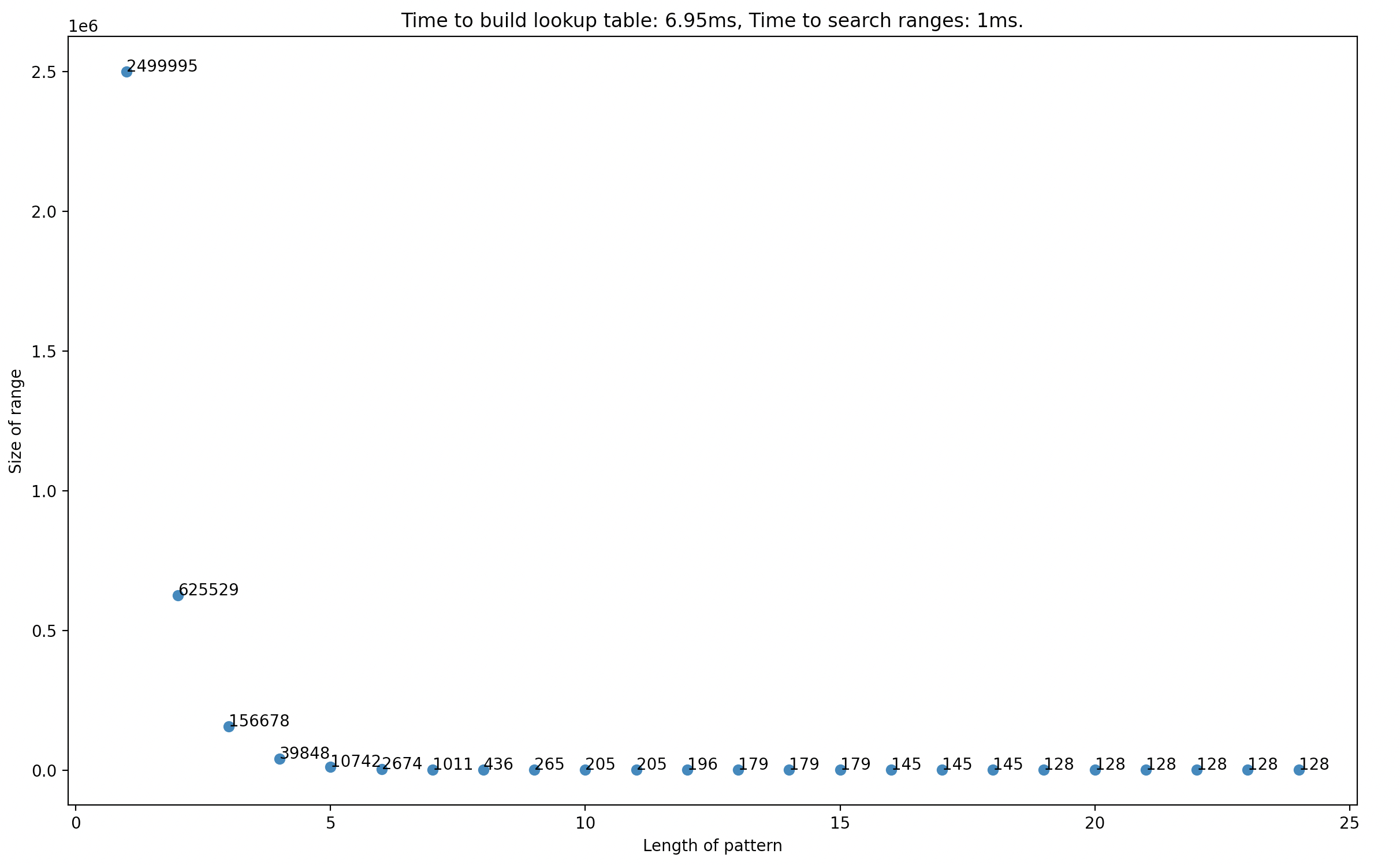}
  \caption{Change of search range as pattern length increases}
  \label{fig:range}
\end{figure}

In the implementation, we used a suffix array created by BurntSushi \cite{BurntSushi} which creates a suffix array in $O(n)$ time using $O(n)$ space using Farach's Algorithm \cite{Farach}. We then implemented the outlined algorithm above along with the creation and query algorithms for the four lookup table methods using the Rust language. In theory, when pre-processing time is not a consideration, the Instant lookup table would be the best choice as it has a constant time lookup which is independent of the text length or pattern length. However, in practice, the query times of the other methods are not long enough to merit waiting for a instant lookup table to be built, and the space complexity of an instant lookup table is prohibitive. This result is due to the fact that in a typical application such as a DNA string, we are frequently dealing with querying very small ranges as the partial patterns $p_f$ and $p_r$ searched through the suffix arrays drastically narrow the possible ranges. The change of these ranges are demonstrated over a ten million character text where the average range size of a 25 character pattern is taken (see Figure \ref{fig:range}).

Here, we show the times that it takes to construct different lookup tables. Admittedly, there may be some room for further optimization, but we believe that these are accurate representations of the capabilities of these methods. Notice how much more time the instant lookup table takes to construct. As such, it does not become a realistic algorithm to use and so we do not bench mark it further (see Table \ref{tab:lookup}).
\begin{table}[h!]
    \centering
     \begin{tabular}{| c || c | c | c | c|} 
     \hline
      & Inverse & Root & Log & Instant \\
     \hline\hline
     200 & 61.88µs & 2.13ms & 320ms & 620s\\
     \hline
     5,000 & 335µs & 398ms & 260s &  \\
     \hline
     100,000 & 6.74ms & 151s & & \\
     \hline
     10,000,000 & 750ms & & & \\
     \hline
    \end{tabular}
\caption{Build times for various lookup tables over different input sizes}
\label{tab:lookup}
\end{table}

Given how the other options still perform acceptably over reasonably sized texts, we consider the Inverse, Root, and Log methods and compare their run times over different pattern lengths, running multiple times and over multiple input texts to find the average (see \ref{tab:lookuptimes}).

\begin{table}[h!]
\centering
 \begin{tabular}{| c || c | c | c |} 
 \hline
  & Inverse & Root & Log \\
 \hline\hline
 5 & 61.88µs & 77.67µs & 107µs \\
 \hline
 10 & 335µs & 207µs & 135µs  \\
 \hline
\end{tabular}
\caption{Lookup times for various table methods over a 5000 character DNA string}
\label{tab:lookuptimes}
\end{table}

\begin{table}[h!]
\centering
 \begin{tabular}{| c || c | c |} 
 \hline
  & Inverse & Root \\
 \hline\hline
 5 & 61.88µs & 449µs \\
 \hline
 10 & 102µs & 178µs \\
 \hline
 25 & 308µs & 406µs \\
 \hline
 50 & 833µs & 934µs \\
 \hline
\end{tabular}
\caption{Lookup times for various table methods over a 100,000 character DNA string}
\label{tab:lookuptimes2}
\end{table}

We see from this analysis that the inverse lookup table method far outperforms the other methods both in build time (as well as space!), and is comparable with the Root and Log methods in run time. While this seems counter intuitive as the worst case time complexity is $O(n)$, in reality the ranges that are being searched are very small in comparison to the text size. The root table method fails to be as efficient as it could be, given that the ranges are frequently smaller than the block size, meaning that searches are always happening using the inverse method as opposed to leveraging the advantage of the pre-computed blocks. 

\subsection{Other Considerations}
\label{sect: Other Considerations}
There are some interesting theoretical edge cases to consider, specifically if the input text is very self similar, which would lead to large query ranges. To demonstrate this, we query over a text of five thousand $A$'s to demonstrate where the other methods become useful.

\begin{table}[h!]
\centering
    \begin{tabular}{| c || c | c | c | }
    \hline
    & Inverse & Root & Log \\
    \hline
    \hline
    Time & 11.6ms & 2.96ms & 633µs \\
    \hline
    \end{tabular}
    \caption{Lookup times across a 5,000 character identical text}
\end{table}

Note that the Log method becomes very time efficient here while the Inverse method starts to falter as each pair of ranges is virtually the same size. The Root method is a compromise between the two, but only because of the auxiliary inverse tables it is equipped with. Interestingly, without the inverse arrays, the root table method does not perform well. This result is due to the fact that it is idealized for when the text fits perfectly into a set of blocks. In this particular experiment, this was not the case (and almost never is), so many extra queries outside the blocks have to be performed. This results in many extra, and slow, comparisons that the Inverse method does not have to make as it simply needs to check if the index is within a range. In contrast, the Root method needs to compare each value outside the blocks against the full range in the other suffix array, meaning that many more comparisons take place.

While the above edge case is theoretically interesting, in most real-world applications, the ranges will remain small and so the inverse table method comes out on top as the most time and space efficient method in practice. While it admittedly has a worst case complexity of $O(n)$, the suffix array query ensures that the range will almost never be that large, resulting in much faster queries. To show the power of the inverse table method, we search for some very large patterns within a 100 million character text file. To pre-process an input this large, it takes about 4.5 minutes per suffix tree and an additional 12 seconds to build the Inverse lookup tables.

\begin{table}[h!]\renewcommand{\tabcolsep}{4pt}
\begin{tabular}{| c | c | c | c | c | c | c | c | c | c | c | c |}
\hline
10 & 25 & 50 & 100 & 250 & 500 & 1000 & 2500 & 5000 & 10k & 50k & 100k \\
\hline
19.0ms & 3.99ms & 2.91ms & 6.50ms & 22.4ms & 65.6ms & 220ms & 1.19s & 4.53s & 17.65s & 430s & 1730s\\
\hline
\end{tabular}
\caption{Lookup times for increasing pattern lengths}
\end{table}

Here we see that over very large patterns, search time does start to slow down, especially as the input text $T$ increases in size as well. However, given the advantage of speed when building the inverse table, we still believe that this method remains the most practical method of choice.
\newpage

\bibliographystyle{plain}
\bibliography{main}
\end{document}